\documentclass[a4paper,english]{article}

\usepackage{arxiv}
\usepackage{hhline}
\usepackage[LGR,T1]{fontenc}
\usepackage[latin9]{inputenc}
\usepackage{babel}
\usepackage{array}
\usepackage{float}
\usepackage{caption}
\usepackage{url}
\usepackage{graphicx}
\usepackage{esint}

\usepackage[authoryear]{natbib}
\usepackage[pdftex,unicode=true]
 {hyperref}

\makeatletter

\pdfpageheight\paperheight
\pdfpagewidth\paperwidth

\ProvideTextCommand{\~}{LGR}[1]{\char126#1}

\providecommand{\tabularnewline}{\\}

\newenvironment{lyxlist}[1]
	{\begin{list}{}
		{\settowidth{\labelwidth}{#1}
		 \setlength{\leftmargin}{\labelwidth}
		 \addtolength{\leftmargin}{\labelsep}
		 }}
	{\end{list}}

\usepackage{hyperref}

\hypersetup{
    colorlinks=true,
    linkcolor=blue,
	citecolor = magenta,
    filecolor=magenta,      
    urlcolor=cyan,
}

\makeatother

\begin{document}
\title{How has labour market power evolved?\\ {\large{}Comparing labour market monopsony in Peru and the United States}}
\author{Jorge Davalos\\ Universidad del Pacif\'{i}co \\ Department of Management \\ Lima, Peru \\ \url{je.davalosc@up.edu.pe}
 \And
Ekkehard Ernst \\ International Labour Organization \\ Research Department \\ Geneva, Switzerland \\ \url{ernste@ilo.org}}
\date{31 March 2021\bigskip{}
}
\maketitle
\begin{abstract}
We document and compare the extent and evolution of labour market power by employers on the US and Peruvian labour markets during the 2010s. Making use of a structural estimation model of labour market dynamics, we estimate differences in market power that workers face depending on their sector of activity, their age, sex, location and educational level. In particular, we show that differences in cross-sectional market power are significant and higher than variations over the ten-year time span of our data. In contrast to findings of labour market power in developed countries such as the US, we document significant market power of employers in Peru vis-\`{a}-vis the tertiary educated workforce, regardless of age and sector. In contrast, for the primary educated workforce, market power seems to be high in (private) services and manufacturing. For secondary educated workers, only the mining sector stands out as moderately more monopsonistic than the rest of the labour market. We also show that at least for the 2010s, labour market power declined in Peru. We contrast these findings with similar estimates obtained for the United States where we are able to show that increases in labour market power are particularly acute in certain sectors, such as agriculture and entertainment and recreational services, as well as in specific geographic areas where these sectors are dominant. Moreover, we show that for the US, the labour market power has gradually increased over the past ten years, in line with the general rise in inequality. Importantly, we show that the pervasive gender pay gap cannot be linked to differential market power as men face higher labour market power than women. We also discuss possible reasons for these findings, including for the differences of labour market power across skill levels. Especially, we discuss the reasons for polarization of market power that are specific to sectors and locations.

\medskip{}

\medskip{}

\end{abstract}
\newpage{}
\begin{quote}
``That labor markets have important elements of monopsony power is becoming clear beyond any reasonable doubt.'' \citep{manning20}
\end{quote}

\section{Introduction}

The rise in market power by firms has led to a concern about growing profit mark-ups, spatial inequalities and -- importantly -- increases in monopsonistic wage-setting power on labour markets \citep{imf19}. To date, however, only the evolution of mark-ups in selected OECD countries has been documented and analysed. In contrast, no encompassing study exist as to (i) the extent to which monopsonistic wage-setting power has increased across a broad range of developed and emerging countries, (ii) the extent to which such an increase would have contributed to a rise in within-country inequalities, and (iii) the importance this change in labour market power might have had in explaining global trends in labour income share and inequality.

The rise of within-country inequality and the simultaneous fall in the labour income share in a large number of (mostly) advanced economies is well documented \citep{milanovic16}. What is less clear are the origins of this inequality and whether common trends across countries indicate common underlying factors. Several underlying factors have been highlighted as potential drivers in inequality, including the rise of super-star firms in the digital economy \citep{autor20,gutierrez20,covarrubias19}, the existence of firm-specific, possibly intangible assets \citep{caballero07,deridder19}, globlalization \citep{autor13b}, changes in the tax structure \citep{ciminelli19}, the erosion of labour market institutions \citep{jaumotte15}, a declining degree of market competition due to rising market entry barriers (e.g. non-compete clauses, occupational licensing\footnote{See the discussion in \url{https://samueldodini.com/files/Dodini_occupation_licensing_9_21_2020.pdf} and the literature cited therein}, land-use regulation, see \citealp{cunningham19}) and low labour market mobility (e.g. due to housing sector constraints, \citealp{ernst18}). Some but not all of these explanations would point towards higher market power for firms also on the labour market. Others suggest that market power might increase only for a sub-group of firms on the labour market. For instance, the rise of super-star firms might increase labour market power for only a small group of companies (those which are highly attractive for workers to join) but market competition for talent might remain the same or become even worse for the rest, leading to rising wages among the much larger tail of the distribution of firms. 

The link between inequality and rising labour market power has attracted significant research over the past decade. Under the Obama administration, the Council of Economic Advisors devoted an entire study on the phenomenon of labour market monopsony.\footnote{\href{https://obamawhitehouse.archives.gov/sites/default/files/page/files/20161025_monopsony_labor_mrkt_cea.pdf}{US government discussion of labor market monopsony}} To date, most studies in this area focus on the United States and on particular parts of the labour market therein, such as the market for online and platform work.\footnote{\url{https://voxeu.org/article/monopsony-online-labour-markets}} What is missing is a more encompassing study covering a large part of the labour markets across different sectors, occupational groups, by gender and by age. Also, even though a rise in income inequality is observed both in advanced and emerging countries, only few studies exist that would look at labour market monopsony in other, non-OECD countries.\footnote{\citet{sokolova21} provide a meta-study of existing estimates across a range of estimation methods and countries.} Finally, existing studies typically focus on a cross-sectional analysis, looking at one point in time rather than the evolution over longer periods, which makes it hard to link the \textit{rise} in inequality to a \emph{rise} in labour market monopsony.

In the following, we make use of a structural estimation of the labour market to assess the degree of labour market monopsony, following a methodology originally suggested by \citet{ridder98} and \citet{ridder03}. This approach allows us to provide detailed estimates of labour market power for different groups on the US and the Peruvian labour market for the different years from 2010 to 2020. We discuss and compare various estimation strategies to arrive at a labour market monopsony indicator using standard labour force survey information and analyse our preferred estimation strategy. Our approach to estimate labour market concentration contrasts with current ones as we make use of a structural model that allows us to simultaneously estimate the degree of monopsony for different labour market groups, by sector, age and educational level, providing a more detailed overview than approaches that focus on firm-level concentration (e.g. \citealp{amodio20}). Moreover, using the rich labour force survey information available for the US and Peru we are able to track and compare the evolution of labour market power for both countries.

Our estimates document significant cross-sectoral, occupational and age differences of the degree of labour market monopsony. In particular, we show that contrary to intuition, in the context of the Peruvian labour market it is the high-skilled group that faces more severe labour market monopsony. Also, labour market monopsony is higher in well-paying occupations such as manufacturing and government services compared to agriculture. Finally, we document that labour market monopsony has declined across the board in the second half of the last decade in Peru but not in the United States.

The paper is structured as follows. The next section provides an overview of the literature on labour market monopsony, which has received renewed interest over recent years, in particular with the rise of non-standard forms of employment via online labour platforms. We discuss and compare the estimation strategies used by different authors and highlight that most of the recent literature has focused on various measures of labour and product market concentration to obtain a measure of labour market monopsony. In contrast, our approach relies on a structural estimation of the labour market, avoiding the porblem of identifying the relevant market, including possible cross-market spillovers. In section 3, we discuss our data and methods, including an overview of different estimation strategies for our structural model approach. Section 4 summarises the main findings of our estimations and discusses the implications. A final section concludes and discusses possible policy choices given the heterogeneous degree of market power that we document in both countries. 

\section{What do we know about labour market power?}

The discussion on the importance of market power on labour markets has received increased attention over the past decade. While the theoretical concept of labour market monopsony is well known to economists, most empirical and policy analysis of the labour market still focuses on the competitive market assumption or -{}- at best -- on a bilateral monopoly with wage bargaining as in the by now standard labour market search and matching model \citep{pissarides94,pissarides00,MoPi94}. As discussed by \citet{manning03}, until recently most economists found it unlikely for employers to exercise market power other than in small pockets and under specific circumstances. Partly, this assessment was based on particular circumstances of minimum wage adjustment an the following labour market reaction, which was not thought in line with theoretical predictions (see, for instance, \citealp{neumark21}, and the examples studied therein). Nevertheless, a substantial body of research has developed in documenting the (country) circumstances and types of jobs for which monopsony could be empirically established. Before presenting our own method and results, we therefore delve into an overview of the existing evidence in the following.

\subsection{What is labour market monopsony and why do we care?}

Labour market monopsony was first described and discussed by Joan Robinson \citeyearpar{robinson69} and refers to a situation where (a large group of) job seekers face a single or few employers on the labour market, the latter situation also being described as oligopsony. In practice, a single firm demanding labour might seem to be rare but different configurations can arise -- for instance when spatial mobility is limited, few big companies collude or job seekers have very specific profiles for which only few jobs exist -- effectively limiting competition between potential employers for workers.

In practice, any situation in which the elasticity of (labour) supply is not infinity -- and hence shifts in the labour demand curve affect the wage a company has to pay -- allows a company to excercise some power in setting the wage rather than being a wage taker as would be the case under perfect competition. This is true even in the case when short- and long-run elasticities of labour supply differ and where the long-run labour supply would be perfectly elastic \citep[p. 94]{boal97}. Such inelastic supply can arise for several reasons, other than one single firm - or a small number of colluding ones - offering jobs. For instance, when (geographic) mobility is costly or (legally) restricted, hiring costs increase with wages. The latter occurs, for instance, through occupational licensing, which - while protection specific occupations - has adverse effects on the market power of workers in occupations with similar skill and task profiles \citep{dodini20}. Similarly, when skills and competencies are only imperfectly substitutable, labour services become differentiated and workers face monopolistic competition for their offer \citep{card18}. In such a situation, dubbed ``dynamic monopsony'' by \citet{ashenfelter10} as workers cannot switch employers costlessly to exploit wage differentials, both the separation rates amd recruitement probabilities depend on wages and firms can exploit an inelastic labour supply \citep{manning03,manning20}.\footnote{The dynamics here come from the consideration of worker flows, not from any difference between long- and short-term labour supply elasticities. } Importantly, such situations of imperfect substitutability might arise more frequently as workers acquire more job-specific skills that cannot easily be used elsewhere \citep{bachmann20}.\footnote{``We find that workers performing mostly non-routine cognitive tasks are exposed to a higher degree of monopsony power than workers performing routine or non-routine manual tasks Job-specific human capital and non-pecuniary job characteristics are the most likely explanations for this result''} Finally, jobs might not only be differentiated according to the skills they require but also regarding the amenities and perks employers offer to their workers and which might appeal to them differently because of preference heterogeneity (see \citealp{manning20}, for a discussion and synthesis of these different approaches).

Labour market monopsony - similar to other forms of deviations from perfectly competitive markets - have significant implications for both efficiency and equity considerations, some of which we will present here. First, in the standard form of a static monopsony, firms typically tend to offer too little employment at too low wages in comparison to perfect competition (\citealp[p. 88]{boal97}; \citealp{benmelech20}). Partly, this is driven by the lack of outside opportunities for job holders, which depresses their wages \citep{bassanini19}.\footnote{\url{https://www.legos.dauphine.fr/fileadmin/mediatheque/recherche_et_valo/LEGOS/cv/publications/caroli_2020.pdf}} A traditional answer to such inefficiencies is the introduction of a minimum wage that allows to lift wages, thereby helping expand labour supply and increasing employment. However, such conclusions need not hold when considering search frictions as source of market power, and employment might actually be inefficienctly high rather than too low, depending on the supply elasticities of firms and workers. Most importantly, search frictions can create thick-market effects with multiple (efficient) equilibria that can be pareto-ranked \citep[ch. 3]{manning03}. Monopsonistic competition might also contribute to gender discrimination for which the higher propensity for women to be hired out of inactivity might be one indication. Finally, monopsonistic power tends to reduce firm investment in (general) training; as workers do not receive their marginal product, they will also not have an incentive to invest in their human capital to the (socially) optimal level, leaving the market with less human capital than would socially be optimal.

Labour market monopsony and concentration has also adverse effect at the macro-economic level. First, rising market concentration is associated with an increase in wage inequality \citep{cortes20} and a fall in the labour income share \citep{berger19}. As increasing labour market concentrations reduces market pressure, more inefficient firms survive in the market for longer, depressing labour productivity (\citealp{imf19}; \citealp{deloecker17}; \citealp{banerjee20}; \citealp{Jiang2017}).\footnote{To the extent that market concentration is driven by a rise in intangibles as suggested by \citet{deridder19}, such a link is more complex: An initial increase of market concentration could follow the entry of highly innovative companies that displace incumbents and build up market shares quickly due to an increasing accumulation of intangibles. Once a certain market dominant position has been reached, innovative activity will fall again and further increases in market concentrations would be linked to a deceleration in innovation and productivity. } Finally, with rising market concentration and power, firms can shield themselves better against shocks, lowering the pass-through of input prices and reducing the effectiveness of monetary policy \citep{heise20}.

\subsection{How to measure labour market concentration}

Three main approaches have been used to measure the extent of monopsonistic power by firms. A first strategy consists in measuring firm concentration, for instance through an analysis of market share of the largest companies in a particular country, sector or local labour market. Such an approach is particularly useful when labour market information is scarce and does not allow a more direct approach. \citet{amodio20}, for instance, calculate market concentration for 204 local labour markets in Peru and estimate a strong negative impact of firm level concentration on wages, both for workers in informal and formal employment. Similarly, \citet{manning10} uses average plant size estimates of local labour markets (commuting zones) to establish the impact of firm concentration on wage premia. In particular, he finds that higher plant sizes in agglomerations allow for labour market power to depress wage premia in comparison to competitive wages in order to compensate for congestion effects. 

A second approach consists in estimating the elasticity of the (firm-specific) labour supply. Indeed, when more detailed information is available, estimating the elasticity of labour supply a firm faces allows to properly assess the extend to which it can exercise monopsonistic power. As mentionned in the previous section, a simple concentration of firms in a particular segment might not be sufficient to precisely measure labour market power. Comparing labour market concentration and labour supply elasticity \citet{azar19b} confirm that both approaches - labour market concentration and labour supply elasticity - are inversely related to each other and hence informative about labour market power. \citet[table 5]{manning11} reviews the available evidence, which refers to situations of quasi-experiments where (public) employers document their hiring difficulties and the reaction of applicants to changes in wages offered. \citet[p. 1003]{manning11} concedes, however, that the resulting labour supply elasticities are too low to be credible, owning to the specific circumstances of the estimates.

Instead, \citet[p.44]{manning03} offers a simple, alternative measure of the degree of monopsony power using the standard search and matching framework: Under the assumption of such a model, hiring out of non-employment rather than through job-to-job transitions increases the power of an employer to fix wages. While this is an appealing approach when comparing measures over time within a given country, cross-country comparisons of such a measure will be of limited information given the institutional differences: Employers in the United States, for instance, tend to lay-off and re-hire workers during downturns much more frequently than their German counter-parts who make use of a specific instrument called ``Kurzarbeit'' that allows firms to suspend wage payments during (temporary) periods of hardship without laying those workers off.

A third approach to estimating labour market power consists in setting up a structural model to simultaneously estimate the demand and supply curves on relevant (local) labour markets. This is the most demanding approach of the three in terms of data requirements. \citet{ridder03}, making use of their estimated equilibrium search model developed in \citet{ridder98}, derive estimates of labour market frictions for a selected number of European countries and the United States. A key challenge of this approach relates to the fact that tenure and worker flows might be determined by characteristics of the job itself. For instance, they might be conditional on a job's wage. Often such information is not readily available, even from micro data (we will get back to this point below). In case, only aggregate information is available, \citet{ridder03} also analyse the unconditional approach, where job characteristics do not influence tenure and worker flows.\footnote{This corresponds to the basic formulation of the search and matching framework, where job separation rates are exogenous.} The unconditional approach despite being less demanding in terms of data is, however, sensitive to the institutional features of the wage setting process and might not yield robust estimates that are comparable over time or across countries.

Finally, an indirect measure can be used by estimating the impact of a change in the minimum wage. For instance, a high degree of monopsony has been used to explain posive employment effects in fast-food restauration after the increase in the New Jersey minimum wage in 1992 as documented by \citet{card95}. Alternatively, in the absence of observing an increase in employment when minimum wages are introduced or raised, researchers have concluded that the degree of monopsony in that specific labour market segment must be low or absent \citep{neumark08,neumark21}.\footnote{See also the discussion here: \url{https://voxeu.org/article/employment-effects-minimum-wages-directions-research}} As noted by \citet{manning03}, however, the observed relationship between the introduction of a minmum wage and changes in employment might depend on the specific market characteristics: Under the assumption of free entry of firms, for instance, an increase in minimum wages can lead to both increases or decreases in employment, even in a situation of monopsony. In other words, the competitive equilibrium and a situation with monopsony would be observationally equivalent even if the introduction of a minimum would still be socially efficient in the latter case.

\subsection{What do we know about labour market concentration?}

In the following we give a brief overview of some estimates of labour market concentration, both across countries and occupations. For a meta analysis of the available empirical analysis across more than 100 studies, we refer to \citet{sokolova21}.

\citet{boal02} summarise historical estimates of labour market power for selected occupations in the United States (see table \ref{tab:MonopHist}). Even though individual observations refer to different time periods and might, therefore, not be comparable, the table shows significant variations across occupations. Importantly, labour market power does not only seem to affect low-skilled workers and depends significantly on contractual clauses as in the case of baseball players subject to reserve clauses.

\begin{table}[htb]
\caption{\label{tab:MonopHist}Degree of monopsony, selected occupations, United
States}

\medskip{}

\begin{centering}
\begin{tabular}{>{\raggedright}m{4cm}>{\centering}m{5cm}>{\centering}m{4cm}}
\hline 
\textbf{Labor market} &
\textbf{Estimated rate of monopsonistic exploitation} &
\textbf{Source}\tabularnewline
\hline 
{\footnotesize{}Baseball players subject to reserve clause} &
{\footnotesize{}$100\%-600\%$} &
{\footnotesize{}Scully (1974), Kahn (2000)}\tabularnewline
{\footnotesize{}Baseball players not subject to reserve clause} &
{\footnotesize{}$\simeq0$} &
{\footnotesize{}Zimbalist (1992)}\tabularnewline
{\footnotesize{}Teachers and nurses} &
{\footnotesize{}$\simeq0$} &
{\footnotesize{}Boal and Ransom (2000), Hirsch and Schumacher (1995)}\tabularnewline
{\footnotesize{}University professors} &
{\footnotesize{}$<5\%-15\%$} &
{\footnotesize{}Ransom (1993)}\tabularnewline
{\footnotesize{}Coal miners in early twentieth century} &
{\footnotesize{}$<5\%$} &
{\footnotesize{}Boal (1995)}\tabularnewline
{\footnotesize{}Textile mill workers in the nineteenth century} &
{\footnotesize{}>0} &
{\footnotesize{}Vedder, Gallaway, and Klingaman (1978), Zevin (1975)}\tabularnewline
{\footnotesize{}Labor market in general} &
{\footnotesize{}$1\%-3\%$} &
{\footnotesize{}Brown and Medoff (1989)}\tabularnewline
\hline 
\end{tabular}
\par\end{centering}
\medskip\parbox{15cm}{\footnotesize Source: \citet{boal02}, the paper and all references cited in the table area available at: \url{http://eh.net/encyclopedia/monopsony-in-american-labor-markets/}}
\end{table}

More recent approaches focus on the analysis of concentration indices and their estimated effect on wage elasticities. For instance \citet{azar20b} estimate concentration indices for local occupational labour markets, using commuting zones as a spatial delineation for local labour markets. Using data from online vacancy postings, they estimate concentration indices for 8000 US local occupational labour markets and document important effects of these concentration indices on posted wages, suggesting a significant degree of labour market power. Using a slightly different data source for online vacancy postings from Burning Glass, \citet{azar20} also document significant differences between labour market vs product market concentration indices, with the former significantly lowering wages. Guest workers in the US have been shown to experience degrees of monopsonistic power that reduces their wages by up to 13\% in comparison to a fully competitive market \citep{gibbons19}. Similarly, for France, \citet{marinescu20} suggest to use a Herfindahl concentration index for local (occupational) labour markets with location identified at the 4-digit occupational breakdown. Using linked employer-employee data, they calculate concentration indices for both firms and workers as two separate sources of monopsony power. In order to identify sector-specific effects of their concentration indices, they run simulations assuming company mergers in individual sectors. In aggregate, the authors document a significant impact of an increase in labour market concentration on employment, but less so of product market concentration. The authors document significant sectoral differences based on their simulation study, with effects largest in retail trade but also building maintenance. Moreover, they demonstrate the existence of industry-spillovers from a rise in sectoral labour market concentration to job loss in other industries. 

Using census and survey data of US manufacturers, \citet{hershbein19} suggest a direct measure of labour market power by analysing the difference between the output elasticity of (plant-level) employment and the labour share in a firm's total cost, i.e. a firm's markdown. In fully competitive markets, this ratio should be one, corresponding to the equality between the marginal product of labour and wages. In contrast, the authors find large deviations with workers earning on average only 65 cents for each dollar produced. Aggregate estimates of markdowns according to this methodology declined from the 1970s to the early 2000s before sharply rising again until 2012, above its level observed at the beginning of the series.

With the rise of labour market platforms (online platforms, gig work), research has focused on how this has affected labour market power. \emph{Prima facies}, labour market platforms are highly competitive as on both sides of the market, buyers and sellers of labour market services bid for the best offer and easily switch. In reality, however, labour market intermediaries fix the conditions besides taking a significant cut for their ``intermediation'' services thanks to incomplete information \citep{kingsley15}. \citet{dube20}, running several experiments on the Amazon Mechanical Turk online labour platform, confirm that labour supply elasticities are, indeed, low. This applies to both recruitment elasticities - hiring workers for specific tasks - and retention elasticities, i.e. keeping workers for additional tasks.

The introduction or (local) variation of minimum wages has also been exploited in other countries to assess the degree of labour market monopsony. In Germany, for instance, a minimum wage was first introduced in 2015. This allows \citet{bachmann17} to use a semi-structural estimation approach with linked employer-employeed data to assess the degree of monopsonistic power in different low-wage industries. They document large sectoral differences, with accommodation and restaurants displaying high employer labour market power whereas little employers in low-wage manufacturing (e.g. food production) have little labour market power. Rather than studying cross-sectoral differences, \citet{azar19} focus on the retail sector specifically, analysing local labour market concentration indices calculated on the basis of online job postings and exploiting state-level variations in minimum wages in the US. They document that more highly concentrated have a significantly lower employment elasticitiy with respect to an increase in minimum wages, with the most highly concentrated local labour markets even having a positive employment elasticity with respect to an increase in the minimum wage.

\section{Data and methods}

\subsection{Data}

\subsubsection{United States}

Labour force information for the United States comes from the Integrated Public Use Microdata Series (IPUMS) of the Current Population Survey (CPS).\footnote{The US micro-data was gathered from IPUMS-CPS web site \url{https://cps.ipums.org/}.} Relevant for our analysis, the US CPS contains information on labour force status, employment status, hours worked, age, sex, education attainment, industry, wage and salary income, region (location) and length of tenure in current job. 

Detailed tenure information is essential for our analysis but is available only ever other year. The information is collected in a supplementary survey in January of that year. We chose to select information over the past decade from the following years: 2010, 2012, 2014, 2016, 2018 and 2020.

To match the underlying assumptions of \cite{ridder03}, only full-time wage workers are selected into our sample i.e. we drop the employed labor force that worked less than 35 hours per week. Wage workers aged between 15 and 64 years are included in our sample.

\subsubsection{Peru}

For Peru, we use the Peruvian National Household Survey (\textit{Encuesta Nacional de Hogares}, ENAHO) for years 2014-2018 as our micro-data source. The \textit{ENAHO} contains information related to  education, health, income, wealth, physical assets and other social aspects of households and individuals. This survey also includes a labor market specific module, the main source of our study. In order to segment the labor market, we retain three key variables from the survey: Economic sector, age (a proxy of experience) and education level. The estimation of the monopsonistic competition index requires two additional variables from the survey: Monthly wages (in Peruvian soles) and the elapsed employment spell (in years). To comply with the methodological framework in \cite{ridder03} , we exclude self-employed individuals and informal salaried workers from our sample.

\subsection{Estimating labour market frictions}

In this subsection we discuss our methodology of estimating labour market power. We make use of the methodological framework developed by \cite{ridder03} but adapt it to the labour market information at our disposal. In the following we discuss our preferred approach. In annex B we also present an alternative approach that is less demanding in terms of the required input data. In annex C we apply this approach to another country, France, and demonstrate its short-comings.

\subsubsection{The conditional inference estimation of labour market frictions}

Our preferred approach builds on an structural model whose equilibrium conditions depend on the distribution of offered $F\left(w\right)$ versus observed $G\left(w\right)$ wages. Under the model assumptions two parameters are key: The arrival rate $\lambda$ (per unit of time) at which workers receive wage offers and the layoffs rate $\delta$ (per unit of time). Thus, the observed wages distribution writes as:
\[
G\left(w\right)=\frac{F\left(w\right)}{1+k\left(1-F\left(w\right)\right)}
\]
where $k\equiv\lambda/\delta$ stands for the average number of job offers in a given spell of employment. This can be interpreted as a labour market friction index with higher values of $k$ implying less frictions. Indeed, as employees have more opportunities to switch with more jobs being offered, the frictions they face on the labour market decline. From the previous condition, the following steady state relationship between the expected elapsed employment spell (left-hand side) and the observed wages distribution arises:
\begin{equation}\label{eq:LMFriction}
\frac{1}{\delta+\lambda\left(1-F\left(w\right)\right)}=\frac{1}{\delta\left(1+k\right)}+\frac{k}{\delta\left(1+k\right)}G\left(w\right)
\end{equation}

where $\delta+\lambda\left(1-F\left(w\right)\right)$ is the hazard rate of the distribution of elapsed job durations. It is also interpreted as job exit rate for workers earning $w$. The reciprocal provides an alternative expression to the hazard rate of elapsed durations:
\begin{equation}\nonumber
\delta+\lambda\left(1-F\left(w\right)\right)=\frac{\delta+\lambda}{1+k\cdot G\left(w\right)}
\end{equation}

\subsubsection{Linear regression (semi-parametric) estimation}

Labour market frictions are identified by the parameter $k$, which can be estimated from the equation (\ref{eq:LMFriction}) above. Its left hand side is interpreted as the average elapsed employment spell at a given wage level. This steady state condition depends on the right hand side determinants and parameters where $G\left(w\right)$ is the cumulative density function of observed wages. Thus, adding an error term to the theoretical relationship yields uni-variate linear model $y=\beta_{0}+\beta_{1}x+u$ such that $y\equiv\frac{1}{\delta+\lambda\left(1-F\left(w\right)\right)}$ and $x\equiv G\left(w\right)$ from which $k$ is easily identified without distributional assumptions by $\beta_{1}/\beta_{0}$.

The caveat of this approach relies on its dependence on observed wage micro-data. In the following, we illustrate this procedure with data for the Peruvian panel of individuals in 2018, from which 12070 elapsed employment spells were observed along with their deflated monthly wages.\footnote{The same approach has been used for US data and all other years.} The sample is the formal salaried labour force aged between 15 and 65 years old. The empirical distribution of job spells and (log) wages is illustrated below:

\begin{figure}[H]
    \centering
    \caption{Distribution of job spells and wages}
    \medskip
    \begin{tabular}{cc}
        \textit{Panel A: Job spells} & \textit{Panel B: Monthly wages}  \\
            \includegraphics[scale=0.5]{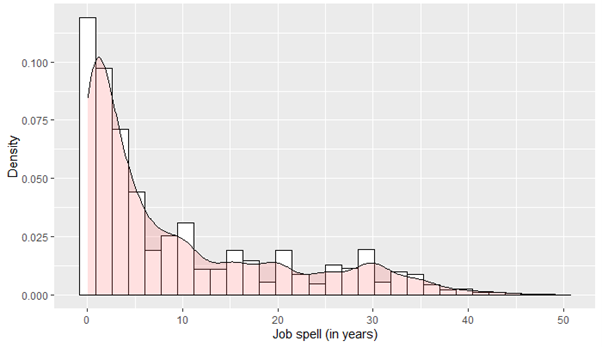} & \includegraphics[scale=0.5]{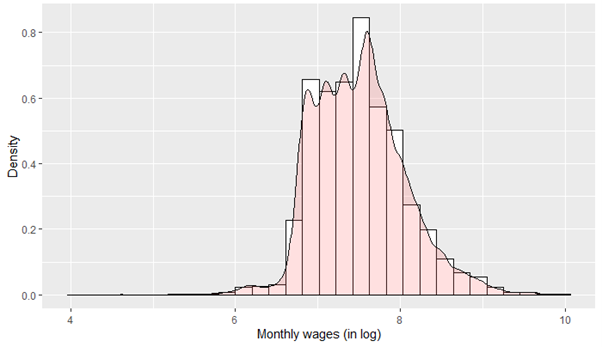} \\ 
            \multicolumn{2}{c}{\textit{Panel C.: Elapsed employment and wages}}\\
            \multicolumn{2}{c}{\includegraphics{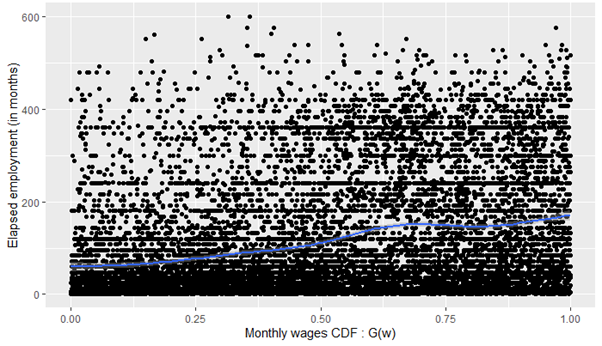}}
    \end{tabular}
    \parbox{14cm}{Source: ENAHO 2018, own calculations}
    \label{fig:peruDistribution}
\end{figure}

As illustrated by the smoothing spline (blue line) in panel C in figure \ref{fig:peruDistribution}, the relationship between elapsed employment and wages can reasonably be approximated by a linear function. Estimating equation \ref{eq:LMFriction}, our index of labour market frictions $k$ can then be retrieved as the ratio of the slope and intercept; its variance can be retrieved by the delta-method:

\begin{table}[H]
    \centering
    \caption{Semi-parametric estimate of labour market frictions $k$}
    \begin{tabular}{lcccc}
 \hhline{=====} & Estimate & S.E. & 2.5\% & 97.5 \% \\ \hline
    k &	2.4220162 & 0.1773811 & 2.0743556 & 2.7696767 \\
    $\delta$ & 0.0687030 & 0.0008862 & 0.0669661 & 0.0704398 \\
    $\lambda$ & 0.1663998 & 0.0108541 & 0.1451262 & 0.1876733 \\ \hhline{=====}
    \end{tabular}
    \label{tab:semiparamK}
\end{table}

\subsubsection{Duration estimation}

Instead of using a semi-parametric approach, we also use a parametric approach on the basis of a duration model, which is vowed to be more efficient (i.e. displays lower variances) as it exploits information on the distribution of employment tenure spells. Elapsed employment duration ($t_e$) may be accurately represented by an exponential distribution: $$ f(t_e) = \theta_x\exp(-\theta_x \,t_e)$$

where $\theta_x$ represents the hazard function. Alternatively, it can be interpreted as the expected number of job interruptions per unit of time or the inverse of the expected elapsed employment duration. $\theta_x$ may be conditioned on a set of explanatory variables ($x$). From \citet{ridder03}\textquoteright s model, the hazard function is defined by the right hand side of equation (2) as a function of the \textit{CDF} of wages ($G(w)$):

$$\theta_x := \frac{\delta+ \lambda}{1+k G(w)}$$

Thus, the likelihood function for the non-censored i-th individual duration is simply defined as $\ln \theta_x -\theta_x \,t_e$. Introducing censoring implies that complete employment spells might not be observed for some observations. For instance, our complete sample of Peruvian workers is censored with respect to a complete employment spell. We could write its likelihood from the probability of having a complete employment spell noted $\tilde{t}_e$ that is longer than the observed elapsed employment duration: $P[\tilde{t}_e > t_e] = 1 - F(\tilde{t}_e)$. In practice, the latter model parameters cannot be identified without a complementary sample of non-censored observations. Combining the  sample of censored ($C$) and non-censored observations ($\bar C$) yields the log-likelihood function :

$$ \sum_{i\,\epsilon \,\bar{C}} [\ln \theta_{x_i} - \theta_{x_i} t_{e_i}]  - \sum_{i\,\epsilon \,C} \theta_{x_i} \, t_{e_i} \equiv \sum_{i\,\epsilon \,\bar{C}} \ln \theta_{x_i}- \sum_{\forall i}  \theta_{x_i} t_{e_i} $$

The structural parameters $\lambda$, $k$ and $\delta$ are implied in $\theta_x$. The following table \ref{tab:paramK} provides an overview of the (maximum likelihood) estimates of the labour market frictions index ($k$) without censoring (0) and under three alternative (arbitrary) censoring levels (40, 20 and 10 years):

\begin{table}[H]
    \centering
    \caption{Parametric estimates at different censoring levels}
    \begin{tabular}{lcccc}
     \hhline{=====} & Estimate & S.E. & 2.5\% & 97.5 \% \\ \hline
      k(0)  & 2.390970 & 0.1159610 & 2.163690 & 2.618249 \\
      k(40) & 2.416392 & 0.1172191 & 2.186646 & 2.646137 \\
      k(20) & 3.679716 & 0.1732006 & 3.340250 & 4.019183 \\
      k(10) & 4.893570 & 0.2387799 & 4.425570 & 5.361570 \\ \hhline{=====}
    \end{tabular}
    \label{tab:paramK}
\end{table}

\subsubsection{Parametric vs semi-parametric estimation}

Even though the parametric relation imposes a distributional assumption, it is a widely accepted one i.e. durations tend to be well represented by exponential distributions as can be seen be the kernel density presented above (Job spell empirical distribution). This implies a gain in efficiency (higher precision). Identification from the linear regression may also be distorted by the non-linear slope of the empirical relationship. In an auxiliary estimation (available upon request) that included informal workers, the parametric estimator proved to be more robust than the semi-parametric one. Including informal workers causes the relationship between elapsed employment and wages to exhibit a higher slope for values of $G(w)$ greater than 0.75. Furthermore, the parametric estimates stay plausible in the presence of informal employment. The parametric estimation is therefore the preferred estimation approach based on the micro-data at hand.

\subsection{Conditional inference of labour market frictions}

Returning to our conditional approach, we now implement the conditional approach on US and Peru\textquoteright s formal salaried labour force from 2014 to 2018 using a linear regression and duration model approach respectively. As mentioned above, the parametric approach is our preferred estimation. Nevertheless, in this subsection we also implement the linear regression while addressing for the potential effects of outlying observations in the data. This is done by means of a robust (to outliers) linear regression which down-weights observations whose residuals are too important.

\subsubsection{Peru: 2014-2018 }

\begin{figure}[H]
\centering
\caption{\label{fig:PeruComparisonEstimates}Comparing semi-parametric and parametric estimates of labour market frictions}
\textit{(Peru: 2014-2018)}

\begin{centering}
\includegraphics[scale=0.5]{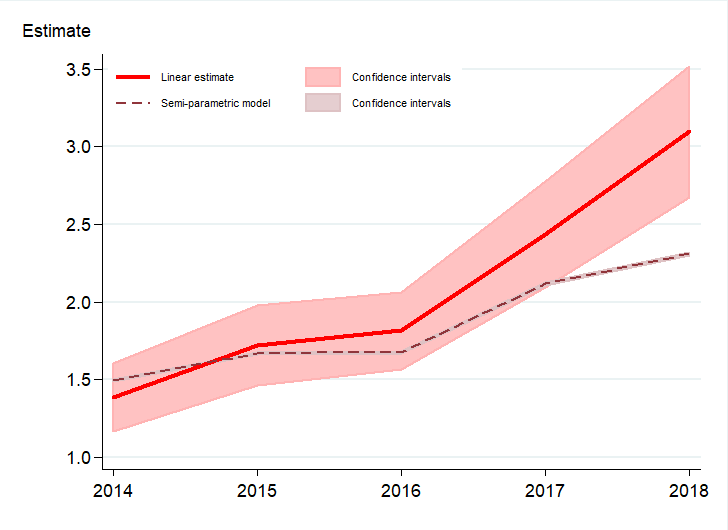}
\par\end{centering}
\parbox{12cm}{\footnotesize Source: ENAHO, own calculations, various years}
\end{figure}

Figure \ref{fig:PeruComparisonEstimates} below compares the robust linear regression and parametric estimates where error bands represent the 95\% confidence intervals. As can be seen, the parametric estimator is more stable across time; the magnitudes are coherent with estimates by \citep{ridder03} for developed countries.

\subsubsection{United States: 2010-2020}

\begin{figure}[H]
\centering
\caption{\label{fig:USComparisonEstimates}Comparing semi-parametric estimates of labour market frictions}
\textit{(United States: 2010-2020)}
\par\bigskip
\hspace{-1.5cm}\begin{centering}
\begin{tabular}{cc}
\textit{Panel A. Linear} & \textit{Panel B. Parametric} \\
\includegraphics[scale=.3]{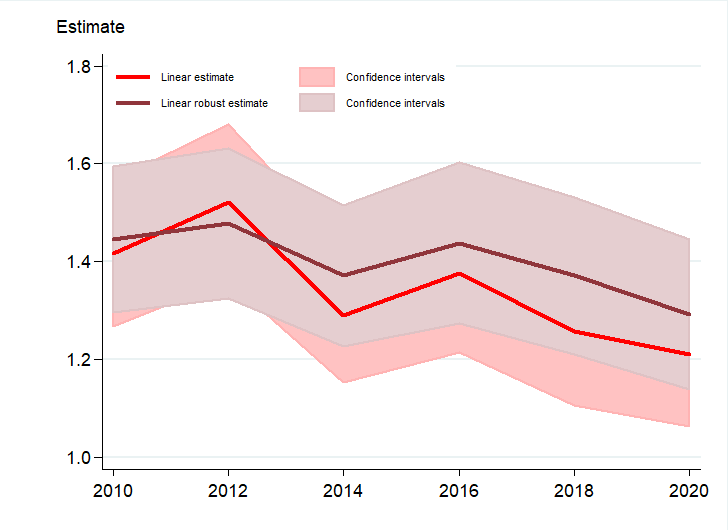} & \includegraphics[scale=.3]{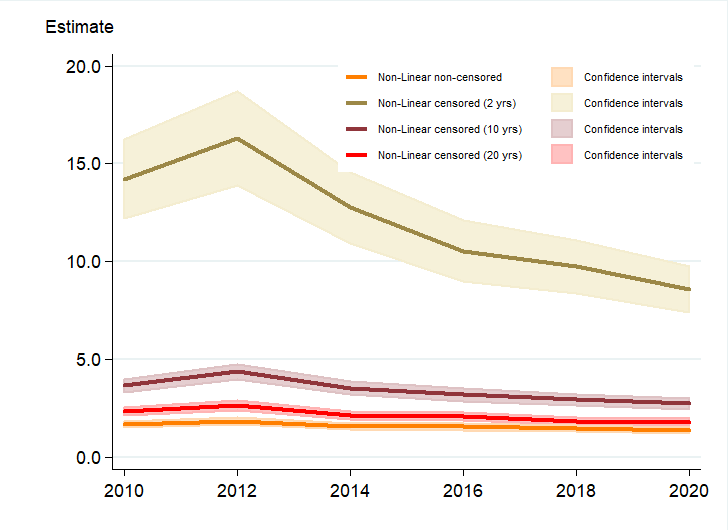}  
\end{tabular}
\par\end{centering}
\parbox{16cm}{\footnotesize Source: US CPS, own calculations}
\end{figure}

Panel A in figure \ref{fig:USComparisonEstimates} reports the estimates of the linear model, both using a standard linear regression \textit{(OLS)} and a robust to outlying observations regression. Panel B reports estimates using the parametric approach for different censoring values (see discussion above). The heterogeneity between the censored estimations suggests that  shorter tenures (likely associated to younger workers) correlate with more frequent job offers during an employment spell. Overall, the results suggest that $k$ exhibits a slight decreasing trend with values from the linear estimation comprised between 1.2 and 1.6., in line with \citet{ridder03}'s earlier estimates for France (1991) that stood at $k=1.4$.

\section{Comparing labour market power in the US and Peru}

Using the conditional estimates of labour market frictions, we can now derive a labour market monopsony indicator for the US and Peru. The advantage of this methodology as discussed in the previous section is that is allows us to calculate labour market power for specific segments of the labour market, taking cross-market effects into account. In our set-up, these segments are defined as on the assumption that labour supply is not expected to switch between segments. Following \citet{ridder98} we define such segments of the labour force according to the following criteria: 
\begin{itemize}
    \item For Peru: Economic sector (CIIU 1-digit, rev.4), potential experience (4 age groups) and education level (primary or less, secondary education, technical, and undergraduate or above);
    \item For the United States: Industry, educational level (2 groups), region and gender.
\end{itemize}
To avoid unreliable estimates due to the small size segments, only
those with at least 30 observation are reported (around 60 segments).
From \citet{ridder98}, the monopsonistic competition indicator for
a given segment is defined as:
\[
\mu=\frac{E\left(w\right)-E\left(\underline{w}\right)}{\left(1+k\right)E\left(w\right)-E\left(\underline{w}\right)}
\]
where $E\left(w\right)$ and $E\left(\underline{w}\right)$ are the expected wage and wage distribution lower bound respectively. Compliance with labour regulations for low productivity workers suggest that $E\left(\underline{w}\right)$ should be given by the institutional minimum wage whereas for high productivity workers $E\left(\underline{w}\right)$ may best be described by a reservation wage that is likely to be higher than the institutional minimum wage. Moreover, in the case of Peru, compliance with its minimum wage legislation is likely to be low and, consequently, the threshold provided by the (legal) minimum wage is not binding.

In their approach \citet{ridder98} provide a structural model that allows the estimation of $E\left(\underline{w}\right)$. However, such estimation is vulnerable to the structure imposed to the data and data measurement errors. For the sake of parsimony, we estimate $E(\underline{w}_j)$ as the wage distribution lower bound given by a \textit{boxplot} minimum estimator (lower whisker) well adapted to symmetric distributions. This is calculated from observed wages in log scale which tend to be symmetric. The literature of Extreme Value statistics provides alternatives estimators that could be explored further.

The mechanisms that determine the lower bound $E\left(\underline{w}\right)$ of the wage distribution are expected to be heterogeneous across countries depending on labour market compliance and productivity levels. Our  implementation makes no prior assumptions regarding institutional factors that determine certain segments minimum wages.

\subsection{The evolution of labour market power in Peru}

\begin{figure}[H]
    \caption{Monopsonistic indicator: Peru - 2014-2018}
    \label{fig:Peru_MonopsonyAgg}
    \medskip{}
    \begin{centering}
    \includegraphics[scale=.5]{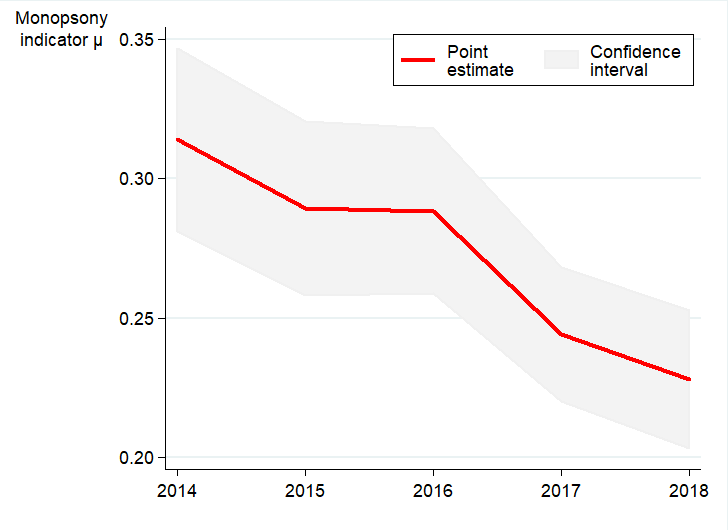}
    \medskip
    \parbox{10cm}{\footnotesize Source: Enaho, own calculations.}
    \par\end{centering}
\end{figure}

For the years 2014 through 2018, figure \ref{fig:Peru_MonopsonyAgg} displays the evolution and standard error of our aggregate monopsony estimate, $\mu$, for Peru, showing a tendency of market power to decline over the previous decade. This is the flipside of the decline in market frictions (or the increase in job offers per employee) indicated in figure \ref{fig:PeruComparisonEstimates} above.

Peru\textquoteright s aggregate estimates of $\mu$ are well above the magnitude found by \citet{ridder03} for OECD countries in the late 1990s. This is somewhat an expected result since Peru\textquoteright s  labour market friction index is lower than the OECD ones reported in \citet{ridder03}, i.e. the number of job offers per employee is much lower. Furthermore, a higher prevalence of labour market monopsonistic power  is consistent with high levels of income inequality in Peru and other Latin-American countries.

The drop in labour market monopsony is also reflected in the aggregate performance of the Peruvian labour market that observed a gradual decline in its informality rate over the 2010s. At the end of the decade, Peru's informality rate had fallen roughly 8 percentage points since 2010, opening up new employment opportunities and faster transition rates to formal employment during that decade. On the other, unemployment rates have been overall very low during the decade (less than 4\%) and most likely did not play an important role in influencing employers' labour market power. On the other hand, the strong macro-economic performance prior to the outbreak of the Covid pandemic in 2020 contributed most likely to strong employment growth in the formal economy, with knock-on effects on labour market power. Indeed, Peru performed well above its regional neighbours during the entire decade of the 2010s (see \citealp{ilo2019}).

\begin{figure}[H]
    \caption{Peru: Monopsony by age and education}
    \medskip
    \begin{centering}
    \textbf{Panel A.: Evolution of monopsony by education} \newline
    \includegraphics[scale=.5]{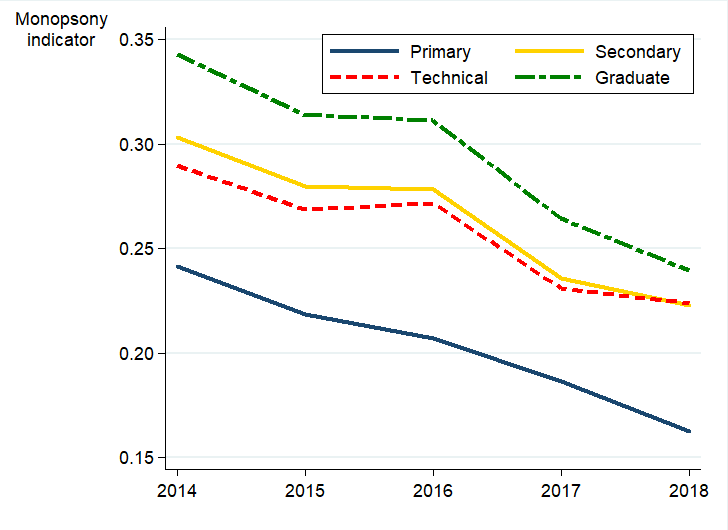} \newline
    \bigskip
    \textbf{Panel B.: Evolution of monopsony by age} \newline
    \includegraphics[scale=.5]{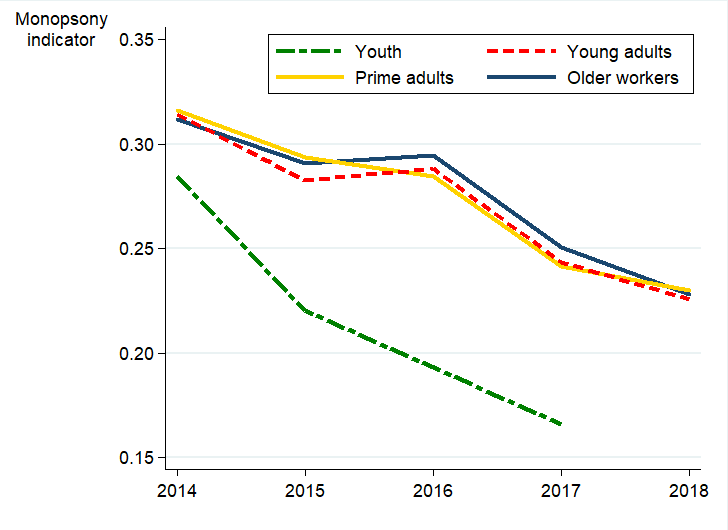}  \newline  
    \parbox{14cm}{\footnotesize Note: Youth=15-20 years; Young adults=21-30; Prime adults=21-45; Older workers=46-65. \\ Source: ENAHO, own calculations}
    \label{fig:Peru_Monopsony}
    \end{centering}
\end{figure}

\begin{figure}[H]
    \caption{Peru: Monopsony indicator - by age, education and sector}
    \medskip{}
    \begin{centering}
    \includegraphics[scale=.5]{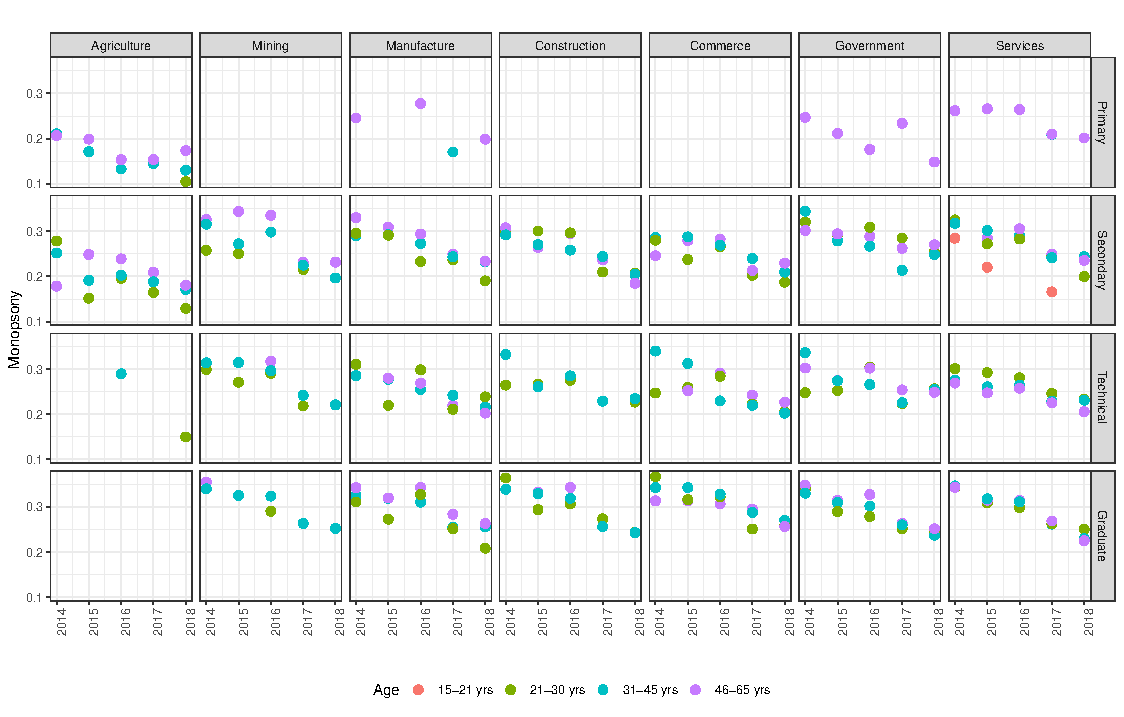}
    \label{fig:Peru_MonopsonyDetail}
    \parbox{14cm}{\footnotesize Note: Youth=15-20 years; Young adults=21-30 years; Prime adults=21-45 years; Older workers=46-65 years. \\ Source: ENAHO, own calculations}
    \par\end{centering}
\end{figure}

At the same time, this fall in labour market monopsony hides significant heterogeneity across different labour market groups (see figures \ref{fig:Peru_Monopsony} and \ref{fig:Peru_MonopsonyDetail}). While the general tendency to a decline in monopsony power is confirmed in all charts across different labour market segments, it seems that younger, less well educated (formal sector) workers in agriculture tend to experience lower monopsonistic wage setting behaviour than graduates in services sector, irrespective of whether services are private or public (government).

Young workers, in particular below the age of 21, seem to benefit from more opportunities to switch jobs, which lowers employers' market power. The same goes for workers with less than secondary education (i.e. "primary" education), most likely because their outside option in the informal economy is larger than for workers with technical or university ("graduate") education. 

This finding fits well with the general characterisation of Peru's labour market as being "multi-segmented", a feature that it shares with many other Latin American countries. Indeed, besides a lower segment related to subsistence activities, many workers and companies prefer to switch to the informal labour market in order to avoid taxation and regulation \citep{maloney04}. In such a labour market, low-skilled (and young) workers that have opportunities in informal employment typically face low switching costs and can command a higher share of their otherwise still meager salary.

In well paying sectors, however, such as mining or manufacturing such outside options are typically not available. Consequently, labour market power by employers is much higher in these sectors, regardless of age and experience than when switching costs are low (see figure \ref{fig:Peru_MonopsonyDetail}). Moreover, figure \ref{fig:Peru_MonopsonyDetail} also demonstrates the significant barriers to (sectoral) mobility. Upward mobility into better paying sectors such as mining or (formal) construction work is not being observed in our sample for workers with less than secondary education (first row). At the same time, no downward mobility is observed for workers with tertiary education ("graduates") into (formal sector) agricultural work. Finally the similar profile of monopsonistic market power for workers above 20 suggest that barriers to mobility are being erected early on in one's professional life and subject to very little variation over worker's career path (see figure \ref{fig:Peru_Monopsony}, panel B). 

A full decomposition of of the contribution of different labour market groups and the historical trend to our labour market monopsony estimate is available in table A.2 in annex A.

\subsection{The evolution of labour market power in the US}

\begin{figure}[H]
    \caption{Monopsonistic indicator: United States - 2010-2020}
    \label{fig:US_MonopsonyAgg}
    \medskip{}
    \begin{centering}
    \includegraphics[scale=.5]{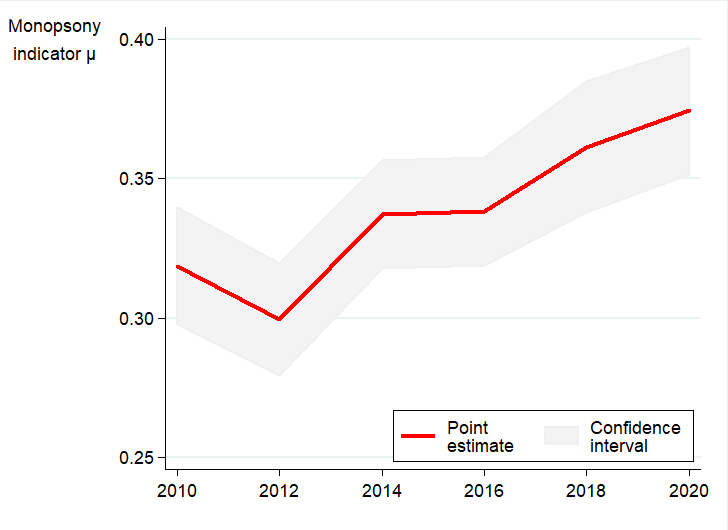}
    \medskip
    \parbox{10cm}{\footnotesize Source: CPS, own calculations.}
    \par\end{centering}
\end{figure}

Figure \ref{fig:US_MonopsonyAgg} displays the aggregate evolution of labour market power in the United States from 2010 through 2020 as well as the confidence interval of our estimates. As discussed, the monopsony indicator $\mu$ can only be calculated every two years because of data limitations concerning job tenure information. After an initial drop at the beginning of the decade - linked to the recovery process after the Global Financial Crisis - our labour market power indicator shows a clear upward trend, the flip side of the increase in labour market frictions documented by our estimates in figure \ref{fig:USComparisonEstimates}. The estimates are relatively precise and indicate that labour market power in 2020 is (statistically) significantly above the level at the beginning of the decade.

This evolution is inline with findings from other works, including regarding the evolution of labour income shares \citep{elsby13, karabarbounis14}, Gini coefficients \citep{horowitz20}, mark-ups \citep{deloecker17} or Herfindahl indices regarding labour market concentration \citep{azar20} such as those presented in section 2. It also points to the fact that the longest but also weakest recovery in post-War US employment was accompanied by rising market power, which might have been instrumental in the relatively slow rate of job creation. Indeed, by early 2020, the employment-to-population ratio still had not recovered to 2007 levels and the unemployment rate needed 8 years to return to pre-recession rates, significantly longer than after previous downturns.\footnote{See, for instance, \url{https://irle.berkeley.edu/the-post-recession-labor-market-an-incomplete-recovery/}}

\begin{figure}[H]
    \caption{Monopsonistic indicator: United States - 2010-2020}
    \label{fig:US_MonopsonyDetail1}
    \medskip{}
    \begin{centering}
    \includegraphics[scale=.65]{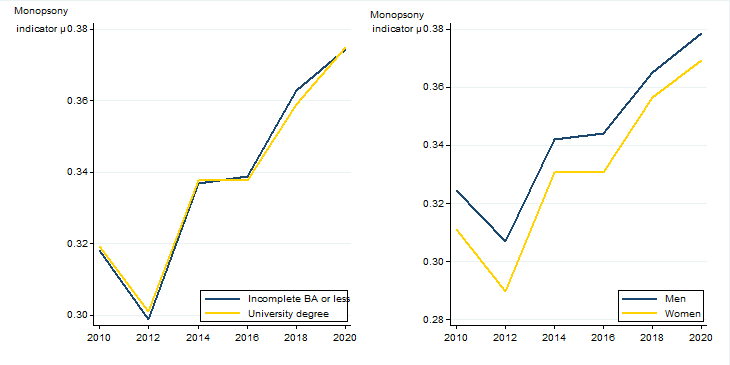}
    \medskip
    \parbox{10cm}{\footnotesize Source: CPS, own calculations.}
    \par\end{centering}
\end{figure}

\begin{figure}[H]
    \caption{Monopsonistic indicator: United States - 2010-2020}
    \label{fig:US_MonopsonyDetail2}
    \medskip{}
    \begin{centering}
    \includegraphics[scale=1]{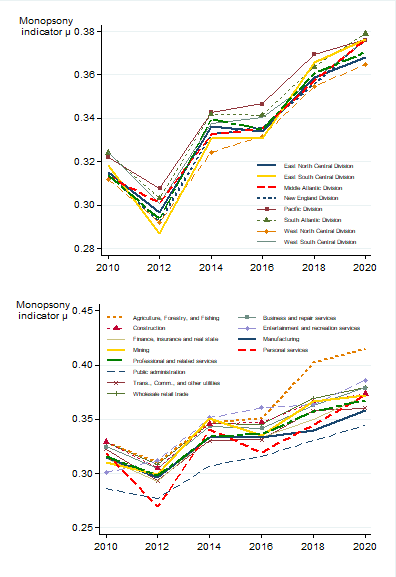}
    \medskip
    \parbox{10cm}{\footnotesize Source: CPS, own calculations.}
    \par\end{centering}
\end{figure}

Figures \ref{fig:US_MonopsonyDetail1} and \ref{fig:US_MonopsonyDetail2} give some further details regarding the sources of these developments. The left panel of figure \ref{fig:US_MonopsonyDetail1}, for instance, shows that the college income and wealth gap discussed by \citet{bartscher20} is most likely not related the differential developments in labour market power. Moreover, in contrast to the situation in Peru discussed above, graduates face a much more fluid labour market that allows them to avoid employers' labour market power. 

The right panel of figure \ref{fig:US_MonopsonyDetail1} documents another interesting finding: Men seem to face higher labour market power than women, most likely because of the sectoral and occupational segregation that leads them to work more frequently in labour market segments with where firms benefit from stronger labour market power. In other words, the large gender pay gap document, for instance, in \citet{ilo18} cannot be explained by differential exposure to labour market power between men and women. Indeed, the report documents that both the raw and the factor-adjusted monthly pay gap between men and women in the United States is around 16\%.

Figure \ref{fig:US_MonopsonyDetail2} breaks down the evolution of labour market power in the US between regions (upper panel) and sectors (lower panel).\footnote{See figure A.1 in appendix A for the geographical breakdown used here.} The upper panel in figure \ref{fig:US_MonopsonyDetail2} demonstrates persistent regional differences in the extent to which American workers are exposed to employers' market power. The cross-regional differences exhibit similarities with the ones documented by \citep{azar20} although our estimator does not permit as fine a geographical break-down as theirs.

The lower panel in \ref{fig:US_MonopsonyDetail2} documents a (large) sectoral heterogeneity in the exposure to labour market power, with some diverging trends in the second half of the observation period. Interesting, the agricultural sector stands out as showing faster increase in labour market power than other sectors. In contrast to the Peruvian experience, public administration shows the lowest degree of labour market monopsony, indicating that US employees have a much higher degree of outside opportunities when working in public administration than their Peruvian counter parts. Nevertheless, labour market power increases there too and there is no sector that shows a divergence from the aggregate trend.

Finally, and in order to better identify how and how much the various labor market segments determine the monopsonistic index level, we decompose its historical mean by means of a regression analysis. The explanatory variables are the ones that define the segments themselves. Thus the coefficients are interpreted as the specific contribution of a given category to the index historical mean. Table A.1 in appendix A provides an overview of our decomposition. The resulting detailed cross-segment evolution of labour market monopsony in the US is illustrated in figure A.2 in appendix A.

\subsection{Country comparisons}

Peru and the United States show striking differences in both the evolution of labour market power and the cross-section differences. This points to significant differences in the way labour markets in both countries work. The US labour market shows a high degree of fluidity with little protection of workers on open-ended contracts. As a consequence, labour market power is felt similarly across different levels of education - and hence skill premia are driven by different factors - and a function of of geographical and sectoral conditions. In contrast, Peru experiences a highly segmented labour market with barriers to entry for informal workers. Interesting, this seems to limit mobility options particularly for high-skilled workers who experience very few opportunities to switch sectors or occupations.

Going back to the earlier estimates for OECD countries established by \citet{ridder03}, reported estimates for $\mu$ typically are below 0.05. In contrast, our estimates for both the United States and Peru are well above such magnitude. This is somewhat an expected result since its labour market friction index is lower than the OECD ones reported in \citet{ridder03}. Furthermore, a higher prevalence of labour market monopsonistic power is consistent with high income inequality trends in Peru and other Latin-American countries. Similarly, the United States experience inequality levels well above other OECD countries, and has seen a stark increase over the 2010s, much faster than some European competitors.

\section{Conclusion and discussion}

The paper analyses and documents trends in labour market power, estimating the degree of monopsony on labour markets in the United States and Peru over the decade between 2010 and 2020. It documents diverging trends between the two countries as well as significant heterogeneity across labour market groups. The sources of heterogeneity differ, however, between the two countries, with large differences in labour market power depending on educational attainment in Peru and stronger cross-sectoral differences in the United States.

A possible extension of our work would be to document similar trends in other OECD countries, notably in Europe. Unfortunately, European labour force surveys do not provide sufficient details on completed job spells that would allow for stable estimates of the wage distribution for job stayers in equation (\ref{eq:LMFriction}). This might require additional assumptions regarding their distributional form, which would have to be validated in each specific case.

Further research could link our estimates for labour market power to a detailed analysis of the implied wage premium. In particular, the fact that some of the trends and cross-segment differences that we document are not well born out in other indicators (college income gap, gender pay gap), might require a careful analysis of the different drivers of inequality between groups within countries. Moreover, the opposing trends observed in Peru in comparison to the US suggest that different countries experience significantly different forces shaping their inequality dynamics, at least as regards labour market power. These are interesting aspects that should be taken up by follow-up research.

\newpage{}

\bibliographystyle{ilostyle}
\bibliography{monopsonyBibTeX}

\section*{Appendix}

\subsection*{A. Tables and charts}

\subsubsection*{Decomposition of labour market power in the US}

\begin{table}[H]
    \centering
    \caption*{Table A.1}
    \begin{tabular}{lcccc}
     \hhline{=====}
    \textbf{Predictor} & \textbf{Coefficient} & \textbf{Standard error} & \textbf{t} & \textbf{p-value}  \\ \hline
    Intercept &	0.348 &	0.003 & 118.04 & 0.000 \\
    Dummy(year=2012) &	-0.021 & 0.002 & -10.62 & 0.000 \\
    Dummy(year=2014) &	0.019 &	0.002 & 9.77 & 0.000 \\
    Dummy(year=2016) &	0.018 &	0.002 &	9.41 & 0.000 \\
    Dummy(year=2018) &	0.041 &	0.002 &	21.79 &	0.000 \\
    Dummy(year=2020) &	0.055 &	0.002 &	28.59 &	0.000 \\
    & & & & \\
    \multicolumn{5}{l}{Education} \\ \hline
    University degree &	0.006 &	0.001 &	4.84 & 0.000 \\
    & & & & \\
    \multicolumn{5}{l}{Demography} \\ \hline
    Gender (women) & -0.013 & 0.001 & -11.40 & 0.000 \\
    & & & & \\
    \multicolumn{5}{l}{Geographic regions} \\ \hline
    East South Central Division &	0.001 &	0.003 &	0.23 & 0.820 \\
    Middle Atlantic Division &	0.001 &	0.002 &	0.41 &	0.685 \\
    Mountain Division &	0.001 &	0.002 &	0.48 &	0.633 \\
    New England Division & -0.001 &	0.002 &	-0.42 &	0.671 \\
    Pacific Division & 0.011 &	0.002 &	4.85 &	0.000 \\
    South Atlantic Division & 0.009 & 0.002 &	4.24 & 0.000 \\
    West north Central Division & -0.004 & 0.002 & -1.87 & 0.062 \\
    West South Central Division & 0.003 & 0.002 &	1.38 & 0.168 \\
   & & & & \\
    \multicolumn{5}{l}{Economic sectors} \\ \hline
     Business and repair services & -0.029 &	0.003 &	-11.03 & 0.000 \\
    Construction & -0.030 &	0.003 &	-9.01 & 0.000 \\
    Entertainment and recreation services & -0.035 & 0.005 & -6.91 & 0.000 \\
    Finance, insurance and real state & -0.034 & 0.003 & -13.04 & 0.000 \\
    Manufacturing &	-0.042 & 0.003 & -15.44 & 0.000 \\
    Mining & -0.038 & 0.006 & -5.87 & 0.000 \\
    Personal services &	-0.037 & 0.004 & -8.77 & 0.000 \\
    Professional and related services &	-0.030 & 0.003 & -11.74 & 0.000 \\
    Public administration &	-0.063 & 0.003 & -24.25 & 0.000 \\
    Trans., Comm., and other utilities & -0.041 & 0.003 & -13.81 & 0.000 \\
    Wholesale retail trade & -0.020 & 0.003 & -8.08 & 0.000 \\
    \hhline{=====}
    \end{tabular}
    \label{tab:US_decompositionLMP}
\end{table}

\subsubsection*{Decomposition of labour market power in Peru}

\begin{table}[H]
    \centering
    \caption*{Table A.2}
    \begin{tabular}{lcccc}
    \hhline{=====}
    \textbf{Predictor} & \textbf{Coefficient} & \textbf{Standard error} & \textbf{t} & \textbf{p-value}  \\ \hline
    Intercept &	0.0833234 & 0.0297318 & 2.802504 & 0.0062646\\
    Trend &	-0.0071257 &	0.0039693 &	-1.795223 & 0.0761310\\
    & & & & \\
    \multicolumn{5}{l}{Education} \\ \hline
    Secondary education &	0.0508233 &	0.0137307 &	3.701434 & 0.0003776\\
    Technical education &	0.0363822 &	0.0139580 &	2.606558 & 0.0107787\\
    Graduate education &	0.0748560 &	0.0137204 &	5.455820 & 0.0000005 \\
    & & & & \\
    \multicolumn{5}{l}{Demography} \\ \hline
    Age(21-30 yrs) &	0.0945188 &	0.0257751 &	3.667053 & 0.0004242\\
    Age(31-45 yrs) &	0.1002579 &	0.0258197 &	3.883002 & 0.0002019\\
    Age(46-65 yrs) &	0.0934590 &	0.0256420 &	3.644762 & 0.0004573\\
    & & & & \\
    \multicolumn{5}{l}{Economic sectors} \\ \hline
    Mining &	0.0667916 &	0.0155412 &	4.297704 & 0.0000453\\
    Manufacture &	0.0469357 & 0.0145092 &	3.234900 & 0.0017276\\
    Construction &	0.0571660 &	0.0144110 &	3.966819 & 0.0001503\\
    Commerce &	0.0563941 &	0.0143073 &	3.941629 & 0.0001643\\
    Government &	0.0802094 &	0.0141042 &	5.686911 & 0.0000002\\
    Services &	0.0687932 &	0.0140284 &	4.903843 & 0.0000044 \\ \hhline{=====}
  \end{tabular}
    \label{tab:Peru_decompositionLMP}
\end{table}

\newpage
\subsubsection*{Regional map of the US}

\begin{figure}[H]
    \centering
    \caption*{Figure A.1: Regional map of the US}
    \rotatebox{90}{\parbox{21cm}{\includegraphics[scale=0.11]{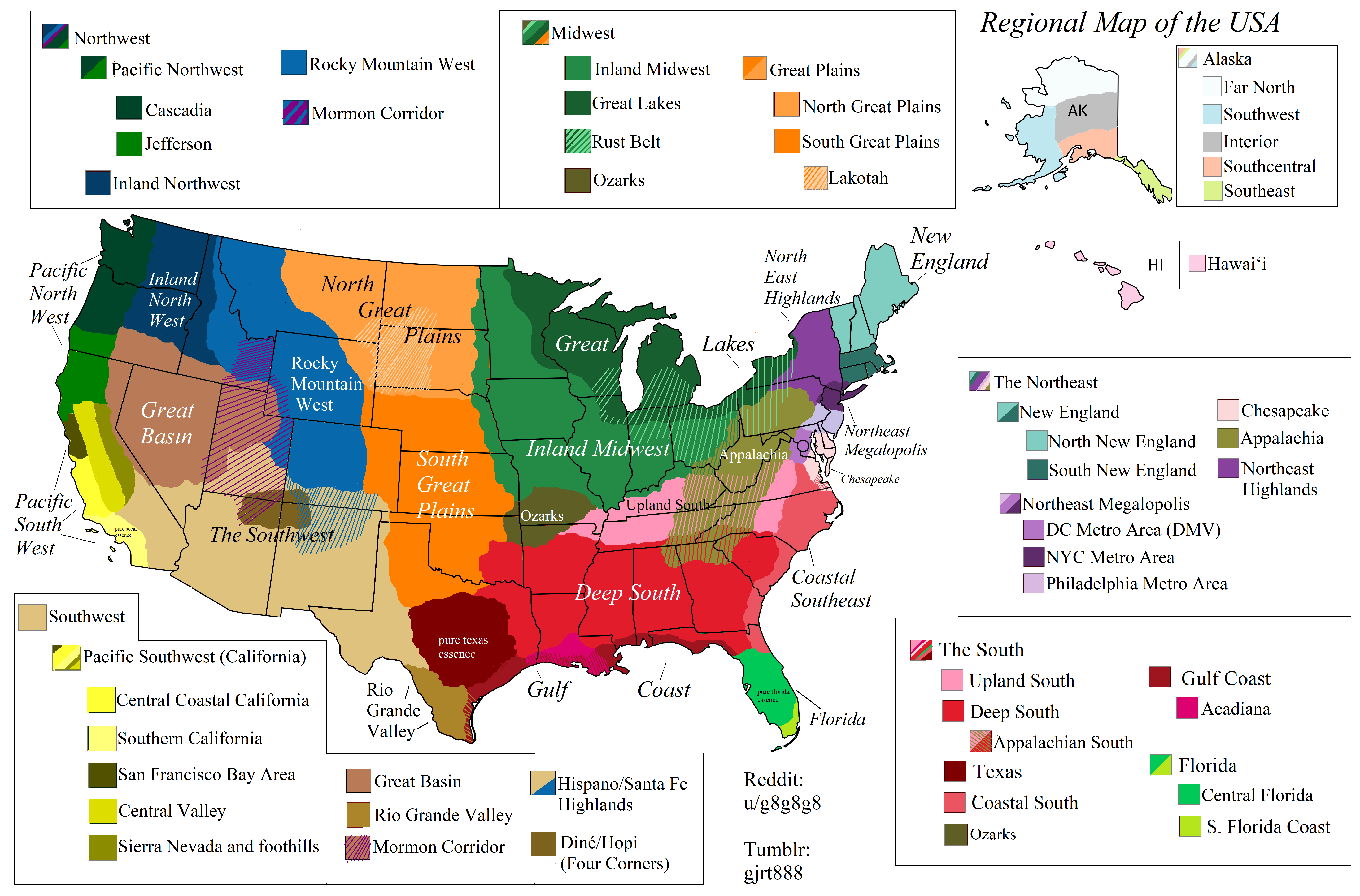} \\ Source: \url{https://i.redd.it/ks05v5y5tqq41.png}}}
    \label{fig:USmap}
\end{figure}

\subsubsection*{Detailed break-down of labour market power in the US}
\begin{figure}[H]
    \label{fig:US_monopsony}
    \caption*{Figure A.2: Monopsony indicator - by age, education and sector: United States}
    \medskip{}
    \begin{centering}
      \rotatebox{90}{\includegraphics[scale=.76]{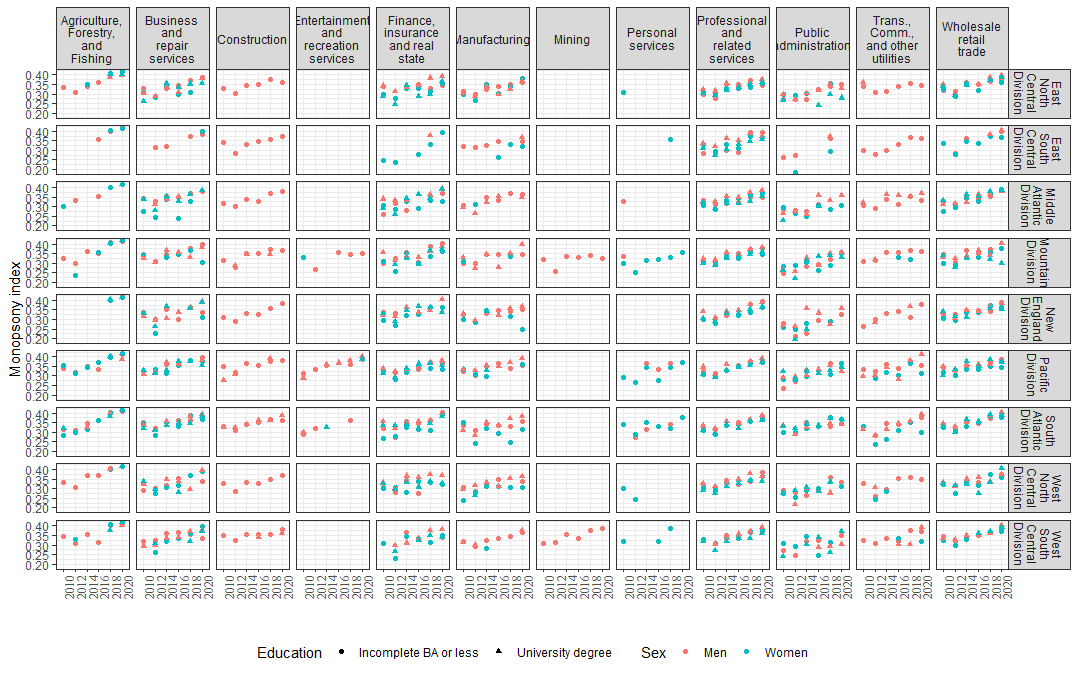}}
    \par\end{centering}
\end{figure}

\newpage\section*{B. The unconditional approach to estimate labour market power}

Given the high demand that the conditional approach imposes on data availability regarding information on job characteristics, \citet{ridder98} develop an alternative theoretical framework for the identification
of the labour friction index based on grouped data. In particular the
authors discuss the identification of \emph{k} from each one of the
sub-populations of the labour force:
\begin{lyxlist}{00.00.0000}
\item [{E-inflow}] : those who transit from unemployment to employment 
\item [{J-inflow}] : those who perform job-to-job transitions 
\item [{E-stock}] : the employed labour force 
\end{lyxlist}
Due to the empirical identification issues associated with an econometric
estimation based on each one of the above mentioned sub-populations,
the authors recommend to ease the identification of $k$ by estimating
$\delta$ (the layoff rate) from an auxiliary model, which we replicate
in the following.

As it turns out, these estimations (from $E$-inflow and $J$-inflow
populations) are highly sensitive to country-specific institutional
setups and measurement errors, although they were expected to be more
efficient. \citet{ridder98} do not test the estimation using the
E-stock. Since we plan to apply these methods to a wider set of countries
wo do not replicate the methods to estimate the $E$-inflows and $J$-inflows
and concentrate on the third (E-stock) which was expected to be less
sensitive to the measurement errors in the data, although less efficient.
Unfortunately, this approach does not provide plausible estimates
when applied to our data for Peru. Nor does it compare favourably
with previous estimates for France.

\subsection*{The auxiliary model for the layoff rate}

The identification of the auxiliary model relies on the observed unemployment
rate, $U$, which can be decomposed into a structural unemployment
rate, $q$, and frictional unemployment:

\[
U=q+\left(1-q\right)\frac{\delta}{\delta+\lambda_{0}}
\]

From it, we define the share of structural unemployment with respect
to total unemployment as $\pi=q/U$. Consequently, the observed distribution
of elapsed unemployment durations $\Psi\left(t_{u}\right)$ can be
expressed as a mixture of structural and frictional unemployment distributions.
This leads the survival function $\overline{\Psi}(t_{u})$ to be written
as a mixture of:
\begin{itemize}
\item 1 (single mass point distribution), this is the probability of being
unemployed after $t_{u}$. This should be one since this is the structural
unemployment probability.
\item and $\exp\left(-\lambda_{0}t_{u}\right)$, the probability of staying
unemployed after tu for those in frictional unemployment:
\end{itemize}
\[
\overline{\Psi}\left(t_{u}\right)=\pi+\left(1-\pi\right)\cdot\exp\left(-\lambda_{0}t_{u}\right)
\]

To operationalize the identification of $\lambda_{0}$ we first aggregate
our elapsed unemployment duration microdata as follows:

\begin{table}[H]
    \centering
    \caption{Aggregating frequencies (ex. Peru - 2018}
    \label{tab:aggrFreq}

    \begin{tabular}{lcccc}
    \hhline{=====} Duration class  & (0,2] & (2,5] & (5,15] & (15,Inf] \\
    Frequency       & 856 & 876 & 199 & 56 \\ \hhline{=====} 
        \end{tabular}
\end{table}

\subsection*{Parametric estimation of the layoff rate}

The likelihood function for the aggregated data shown in the table above must be defined by the probabilities of observing $t_u$ between the upper and lower class ($p_c$) boundaries : $p_c :=\Psi(t_{u_{up}}) -  \Psi(t_{u_{low}})$. Thus the grouped log-likelihood function writes :

$$ \mathcal{\ln L} = n \,\ln (1-\pi) + \sum_{i=2}^{k} n_{i-1} \ln\left[e^{-\lambda_0 t_{u_{i-1}}} - e^{-\lambda_0 t_{u_i}}\right] + n_k \, \ln[\pi + (1-\pi) e^{-\lambda_0 t_{u_k}}]$$

where the $t_{u_i}$ is the i-th class lower bound, $n$ is the total number of observations and $n_j$ is the number of observations at the j-th class. Having an open last class yields the likelihood last term, in it, $t_{u_k}$ represents the lower bound of the last open class. In this particular case, the sum operator index ranges from 2 to $k = 4$ to properly calculate the class probabilities ($p_c$) in our unemployment duration data. 

\subsection*{Grouped E-stock estimation }

In RvdB, alternative estimation strategies based on grouped data are
proposed. The authors mention the posibility of focusing on the $E$-inflow,
$J$-inflow and $E$-stock populations. $E$-inflow and $J$-inflow
consist of workers that transition from unemployment to employment,
and from employment to employement (job-to-job) at a given point in
time respectively. It is argued the employment durations will be described
by different probability distributions. In $E$-inflow and $J$-inflow,
the random variable is the completed job spell. Similarly, the $E$-stock
populations consists of employed workers at a given point in time
whereas the random variable is the elapsed employment spell (not the
completed one). Since aggregated (grouped) data often makes difficult
to disentangle employment spells from $J$-inflow and $E$-inflow
populations, here we develop the estimation approach based on the
$E$-stock sample. It should be noted that RvdB developed $J$-inflow
and $E$-inflow illustrations given the higher efficiency (lower variance)
of the implied parametric estimator. Nevertheless, they also verify
that such estimates are often non-plausible given the measurement
errors i.e. unavailability of pure $E$-inflow or $J$-inflow samples
in grouped data.

The elapsed employment duration $t_{e}$ from a sample of employed
workers at a given point in time has density function:

\[
\psi\left(t_{e}\right)=\frac{\delta\left(1+k\right)}{k}\int_{\delta}^{\delta\left(1+k\right)}\frac{e^{-z\cdot t_{e}}}{z}\cdot dz
\]

From this we derive the \textit{cdf}: $\Psi\left(t_{e}\right)=\int_{0}^{t_{e}}\psi\left(t_{e}\right)dt_{e}$,
which allows us to set up the grouped data log-likelihood as the probabilities
associated to each class in the grouped data:

\[
\ln L=n_{1}\ln\Psi\left(c_{1}\right)+\sum_{j=2}^{J-1}n_{j}\ln\left[\Psi\left(c_{j}\right)-\Psi\left(c_{j-1}\right)\right]+n_{J}\ln\left(1-\Psi\left(c_{J}\right)\right)
\]
where $c_{j}$ represent the lower bound of the $j$-th class interval
in our grouped data, $J$ is the number of groups and $n_{j}$ is
the frequency within the $j$-th category. This estimator is implemented
using the elapsed employment grouped data for Peru:

\begin{figure}[H]
\caption{Elapsed employment duration (in years; Peru, 2018)}

\medskip{}

\begin{center}
\includegraphics[scale=0.6]{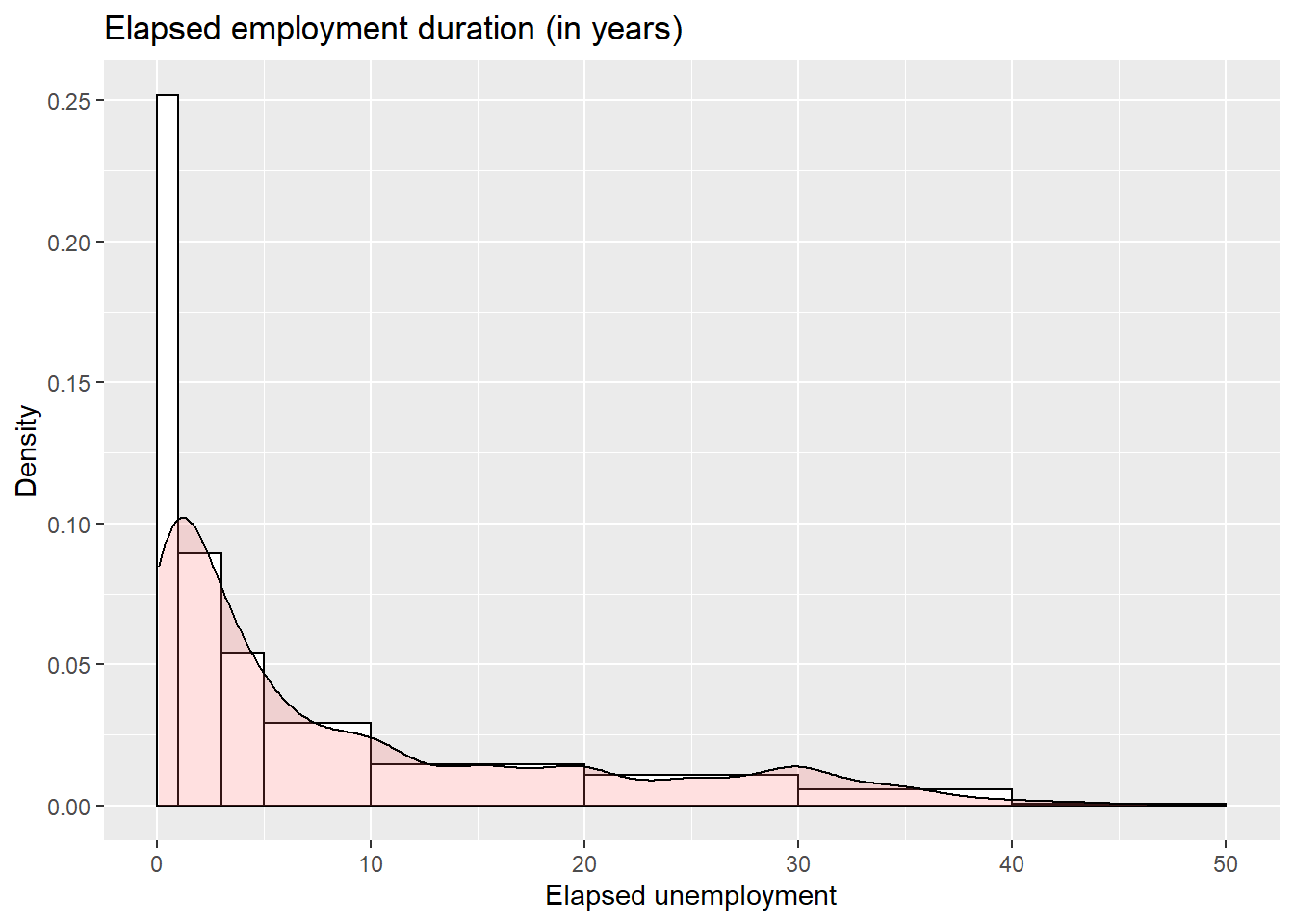}
\par
\bigskip 
\footnotesize
\begin{tabular}{lccccccccc} 
\hhline{==========} Duration  & (0,1{]} & (1,3{]} & (3,5{]} & (5,10{]} & (10,20{]} & (20,30{]} & (30,40{]} & (40,50{]} & (50,Inf{]} \\
Frequency & 3040 & 2155 & 1306 & 1762 & 1759 & 1283 & 692 & 73 &  0 \\ \hhline{==========}
\end{tabular}
\end{center}

\end{figure}

Using these stock and flow estimates allows us to obtain alternative estimates for our labour market friction indicator (see table \ref{tab:semiparamKUn}). As can be seen, the $k$ parameter estimation is implausible which may be explained by its intrinsic high variance (as motivated by \citealp{ridder03}) and measurement (recall) errors for longer spells.

\begin{table}[H]
    \centering
    \caption{Semi-parametric estimates of Peru's labour market frictions $k$}  \textit{(grouped data E-stock approach, 2018)}
    \medskip \\
    \begin{tabular}{lcccc}
 \hhline{=====} & Estimate & S.E. & 2.5\% & 97.5 \% \\ \hline
    k &	2398.1467950 & 6.1766725 & 2386.0407392 & 2410.2528507 \\
    $\delta$ & 0.0575756 & 0.0001542 & 0.0572733 & 0.0578779 \\
    $\lambda$ & 138.0747365 & 0.3695809 & 137.3503713 & 138.7991018 \\ \hhline{=====}
    \end{tabular}
    \label{tab:semiparamKUn}
\end{table}

\section*{\newpage C. Extension for France}

The micro data required for the calculation of our labour market monopsony index are, in principle, available for a large range of countries, including those in the European Labour Force Survey. One drawback that this survey has, however, is that tenure information has been discretized and is not available as continuous information as for the United States and Peru. Typically, only four categories of tenure brackets are available. In the following, we illustrate the adaptation of our procedure for France.

We use France\textquoteright s \textit{Enqu\^{e}te du Travail} (French LFS) where the workers\textquoteright{} tenure has been discretized into four categories has been estimated. This required an adaptation of the previous likelihood functions to consider the grouped nature of the microdata. Although individual wages were available, the conditional approach does not deliver plausible results (see table \ref{tab:France} below). The discretization of the tenure data seems to cause an information loss that penalizes the estimators stability properties. Only 2014 has a 95\% confidence interval that allows for some plausible values. Recall that in the original estimation by \cite{ridder03} France\textquoteright s friction index was about 5 for 1991.

To further check the computational accuracy of our estimator we test it using French data for 1990 and 1991 and compare it with \cite{ridder03}\textquoteright s estimates for the same years. The comparison cannot be exact since the authors produced their estimations using their own unreported aggregations from microdata, while we employ OECD grouped statistics:

\begin{table}[H]

\caption{France: Labour market frictions index}

\medskip{}

\begin{center}
\begin{tabular}{lllll} 
\label{tab:France}
 & Estimate  & SE        & 2.5 \%     & 97.5 \%   \\ \hline
2010 & 0.0000091 & 0.0172834 & -0.0338658 & 0.0338839 \\ 
2012 & 0.0000038 & 0.0100533 & -0.0197003 & 0.0197080 \\ 
2014 & 0.2165852 & 2.0737412 & -3.8478729 & 4.2810433 \\ 
2016 & 0.0000467 & 0.0774138 & -0.1516815 & 0.1517749 \\ 
2018 & 0.0000308 & 0.0449509 & -0.0880715 & 0.0881330 
\end{tabular}
\end{center}
\end{table}

Both the level and time variation of our labour market friction index is substantially higher than either the index produced by \cite{ridder03} and our estimates for the US and Peru reported above. This indicates that the grouped tenure data do not allow a proper application of our estimation methodology to the publicly available French micro data. 

\end{document}